\documentclass[pra,twocolumn,superscriptaddress,amsmath,amsthm]{revtex4}

\usepackage{amssymb}
\usepackage{graphicx}

\newcommand{\bra}[1]{\langle #1 |}
\newcommand{\ket}[1]{| #1 \rangle}
\newcommand{\braket}[2]{\langle #1 | #2 \rangle}
\newcommand{\ketbra}[2]{| #1 \rangle \langle #2 |}
\newcommand{\expect}[1]{\langle #1 \rangle}

\begin{document}
\title{Quantifying non-Gaussianity of quantum-state correlation}
\author{Jiyong Park}
\affiliation{Department of Physics, Texas A\&M University at Qatar, Education City, P.O.Box 23874, Doha, Qatar}
\author{Jaehak Lee}
\affiliation{Department of Physics, Texas A\&M University at Qatar, Education City, P.O.Box 23874, Doha, Qatar}
\author{Se-Wan Ji}
\affiliation{National Security Research Institute, Daejeon, 34044, Korea}
\author{Hyunchul Nha}
\affiliation{Department of Physics, Texas A\&M University at Qatar, Education City, P.O.Box 23874, Doha, Qatar}
\affiliation{School of Computational Sciences, Korea Institute for Advanced Study, Seoul 130-722, Korea}
\date{\today}

\begin{abstract}
We consider how to quantify non-Gaussianity for the correlation of a bipartite quantum state by using various measures such as relative entropy and geometric distances. We first show that an intuitive approach, i.e., subtracting the correlation of a reference Gaussian state from that of a target non-Gaussian state, fails to yield a non-negative measure with monotonicity under local Gaussian channels. Our finding clearly manifests that quantum-state correlations generally have no Gaussian extremality. We therefore propose a different approach by introducing relevantly averaged states to address correlation. This enables us to define a non-Gaussianity measure based on, e.g., the trace-distance and the fidelity, fulfilling all requirements as a measure of non-Gaussian correlation. For the case of the fidelity-based measure, we also present readily computable lower bounds of non-Gaussian correlation.
\end{abstract}

\maketitle

\section{Introduction}
Non-Gaussianity, i.e., deviation from Gaussianity, is a notion of keen interest in many branches of science. In general, a nonlinear process is capable of generating a non-Gaussian distribution, e.g., primordial non-Gaussianity in an inflationary model \cite{Inflation}, fluctuations in a nuclear fusion process \cite{NuclearFusion}, and extremal waves in a nonlinear optical medium \cite{OpticalRogueWave}. Particularly for continuous variable (CV) quantum informatics \cite{Cerf2007}, the competition between Gaussianity and non-Gaussianity has been addressed as a critical issue from fundamental and practical aspects as no-go theorems in a Gaussian regime necessitate non-Gaussian resources. For instance, there is no way to distill Gaussian entanglement by using Gaussian operations \cite{Eisert2002, Fiurasek2002, Giedke2002}, enhance Gaussian-state squeezing by passive Gaussian operations \cite{Kraus2003}, and manifest Gaussian-state nonlocality with Gaussian measurements \cite{Nha2004, Garcia2004}. Furthermore, non-Gaussian resources provide advantages over Gaussian counterparts. Non-Gaussian entanglement can be more robust against Gaussian noises than Gaussian entanglement \cite{Allegra2010, Adesso2011, Sabapathy2011, Lee2011, Allegra2011, Nha2010, Nha2012}. Non-Gaussian states can be distilled by Gaussian operations to increase squeezing \cite{Heersink2006, Suzuki2006, Filip2013} and entanglement \cite{Hage2008, Dong2008}. Non-Gaussian operations can enhance the nonclassical properties, e.g., squeezing \cite{Agarwal1990, SYL2010}, entanglement \cite{Takahashi2010, SYL2011, Navarrete2012, SYL2013, Lee2013, Kurochkin2014, Ulanov2015, Hu2017, SYL2012}, nonlocality \cite{Nha2004, Garcia2004, Park2012}, and multipartite correlation \cite{Kim2013, Nha2007}, as well as the performance of CV quantum informatic tasks, e.g., quantum teleportation \cite{Opatrny2000, Cochrane2002, Olivares2003, DellAnno2007}, quantum dense coding \cite{Kitagawa2005}, and quantum key distribution \cite{Huang2013}.

To rigorously understand the role of non-Gaussianity in quantum science, it is desirable to characterize the non-Gaussianity in a quantitative manner. Non-Gaussianity measures in the quantum regime have been proposed by using geometric \cite{Genoni2007, Genoni2010, Ivan2012, Ghiu2013} and entropic \cite{Genoni2008, Genoni2010} distances. Each measure has characterized the non-Gaussianity of a quantum state reliably employing the distance between a target state and the reference state, i.e., the Gaussian state that has the same first and second moments as the target state. Here we intend to further pursue the study on non-Gaussianity, particularly for correlation, because the non-Gaussianity measures proposed so far are inadequate to characterize the correlation aspect; those measures give a nonzero quantity even for the non-Gaussianity of a product state that clearly has no correlation \cite{Park2, Park22}.

We explore how to quantify the non-Gaussianity of correlation by using various measures, e.g., the relative entropy and geometric distances. We first address a set of required properties for a legitimate non-Gaussianity measure of correlation. We then try to define a measure of non-Gaussian correlation as the difference in the quantum correlation between the target state and the reference Gaussian state, which may seem to be intuitive and reasonable satisfying some relevant properties. In particular, Gaussian extremality has been found in a number of quantum informatic measures. Under the assumption that the covariance matrix of a state is fixed, a Gaussian state maximizes the von Neumann entropy \cite{Holevo1999}
 as well as the quantum information transfer under a Gaussian channel \cite{Holevo2001}. Entanglement measures satisfying superadditivity are minimized by a Gaussian state under the covariance matrix constraint \cite{Wolf2006}. If a quantum mutual information has Gaussian extremality as supposed, it instantly yields a non-Gaussianity measure with a desirable property of non-negativity, which also means that the Gaussian approximation of a given non-Gaussian state does not overestimate correlation. However, contrary to a popular belief in the community \cite{Park1}, we find that the non-Gaussianity measures obtained by subtracting the correlation of the reference Gaussian state from that of the target state is neither non-negative nor monotonic under local Gaussian channels. Our observations indicate that the quantum-state correlations generally do not possess Gaussian extremality. As a remark, our results are unrelated to quantum mutual information in channel theory because the interest of the latter is the information transfer attainable by a given input state through a channel, not the quantum correlation of a given bipartite quantum state (see Fig. 1).
  
We therefore propose a different approach by defining some averaged states relevant to address correlation, which leads us to a legitimate non-Gaussian correlation measure based on, e.g., the trace-distance and the quantum fidelity. We show that these measures satisfy all required properties, particularly being non-negative and nonincreasing under local Gaussian channels. For the case of fidelity-based measure, it is typically hard to obtain an exact value because it requires a nontrivial eigen-decomposition in an infinite-dimensional Hilbert space. Thus, we further provide two reliable and computable lower bounds for the fidelity-based measure. We illustrate the validity of our measures by showing monotonic behavior of the measures under a loss channel.

	\begin{figure}[!t]
		\includegraphics[scale=0.55]{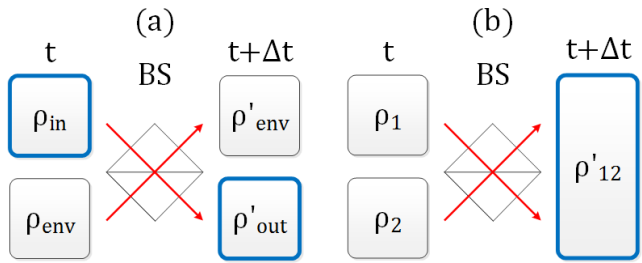}
		\caption{Illustration for distinguishing different notions of quantum mutual information. (a) In quantum channel theory, quantum mutual information is the measure of information transfer through a channel, e.g., a loss channel (the case of $\rho_{\mathrm{env}} = \ket{\mathrm{vac}}$). It quantifies how much information on the input state $\rho_{\mathrm{in}}$ can be retrieved from the output state $\rho_{\mathrm{out}}$. More precisely, it is the amount of shareable quantum information between the sender who encodes the information in $\rho_{\mathrm{in}}$ at time $t$ and the receiver who decodes the information in $\rho_{\mathrm{out}}^{\prime}$ at time $t + \Delta t$. This is quantified as $S(\rho_{\mathrm{in}}) + S(\rho_{\mathrm{out}}^{\prime}) - S(\rho_{\mathrm{env}}^{\prime})$ \cite{Holevo2001}. (b) On the other hand, in this paper, we are interested in the correlation of a bipartite quantum state itself. For example, when two uncorrelated states $\rho_1$ and $\rho_2$ are mixed at a beam splitter, the output state $\rho_{12}^{\prime}$ has the quantum mutual information given by $S(\rho_{1}^{\prime}) + S(\rho_{2}^{\prime}) - S(\rho_{12}^{\prime})$, with the identity $S(\rho_{12}^{\prime})=S(\rho_{1})+ S(\rho_{2})$. Thus, the Gaussian extremality found in panel (a) has no direct relation to the results in our work.}
		\label{fig:confusion}
	\end{figure}

\section{Quantum mutual information} \label{sec:QMI}
The mutual information of a bipartite quantum state is conventionally addressed by
	\begin{align} \label{eq:mi}
		\mathcal{I}_{1} [ \rho_{AB} ] & = S [ \rho_{A} ] + S [ \rho_{B} ] - S [ \rho_{AB} ] \nonumber \\
		& = S [ \rho_{AB} || \rho_{A} \otimes \rho_{B} ],
	\end{align}
where $S [ \rho ] = - \mathrm{tr} [ \rho \ln \rho ]$ is the von Neumann entropy of a state $\rho$ and $S [ \rho || \sigma ] = \mathrm{tr} \rho \ln \rho - \mathrm{tr} \rho \ln \sigma$ is the quantum relative entropy of $\rho$ with respect to $\sigma$. Note that both of the expressions are identical at the level of Shannon entropy. More generally, we may define two types of quantum R{\'e}nyi mutual information in a similar fashion as
	\begin{align}
		\mathcal{I}_{\alpha} [ \rho_{AB} ] & = S_{\alpha} [ \rho_{A} ] + S_{\alpha} [ \rho_{B} ] - S_{\alpha} [ \rho_{AB} ],  \label{eq:rmi1} \\
		\mathcal{I}_{\alpha}^{\prime} [ \rho_{AB} ] & = S_{\alpha} [ \rho_{AB} || \rho_{A} \otimes \rho_{B} ],  \label{eq:rmi2}
	\end{align}
where the quantum R{\'e}nyi entropy of a state $\rho$ is given by $S_{\alpha} [ \rho ] = \frac{1}{1 - \alpha} \ln \mathrm{tr} [ \rho^{\alpha} ]$ and the quantum R{\'e}nyi relative entropy \cite{Muller2013, Wilde2014} of $\rho$ with respect to $\sigma$ is given by
	\begin{equation}
		S_{\alpha} [ \rho || \sigma ] = \frac{1}{\alpha - 1} \ln \{ \mathrm{tr} [ ( \sigma^{\frac{1 - \alpha}{2 \alpha}} \rho \sigma^{\frac{1 - \alpha}{2 \alpha}} )^{\alpha} ] \}.
	\end{equation}

The first type of quantum R{\'e}nyi mutual information in Eq. \eqref{eq:rmi1} is intuitive with an analogy to the intersection of two sets, i.e., $A \cap B = A + B - A \cup B$, and has been employed for measuring information in the Gaussian regime \cite{Adesso2013} as well as for investigating quantum critical systems \cite{Furukawa2009, Singh2011}. However, it can be negative for $\alpha \neq 1$, which is undesirable for quantifying the amount of correlation. We thus adopt the second type of quantum R{\'e}nyi mutual information in Eq. \eqref{eq:rmi2} since the relative entropy is always non-negative for every $\alpha$. It is worth noting that both of the definitions in Eqs. \eqref{eq:rmi1} and \eqref{eq:rmi2} recover Eq.~\eqref{eq:mi} as a limiting case, i.e., $\lim_{\alpha \rightarrow 1} \mathcal{I}_{\alpha} = \lim_{\alpha \rightarrow 1} \mathcal{I}_{\alpha}^{\prime} = \mathcal{I}_{1}$.


In addition, we may define geometrical measures of correlation, namely,
	\begin{align}
		\mathcal{I}_{\mathrm{HS}} [ \rho_{AB} ] & = \sqrt{\mathrm{tr} ( \rho_{AB} - \rho_{A} \otimes \rho_{B} )^{2}}, \label{eq:gmi1} \\
		\mathcal{I}_{\mathrm{TR}} [ \rho_{AB} ] & = \frac{1}{2}\mathrm{tr} | \rho_{AB} - \rho_{A} \otimes \rho_{B} |, \label{eq:gmi}
	\end{align}
where $\mathcal{I}_{\mathrm{HS}}$ and $\mathcal{I}_{\mathrm{TR}}$ employ the Hilbert-Schmidt distance and the trace distance, respectively, to quantify a distance between two states.

The quantum Hilbert-Schmidt mutual information in Eq.~\eqref{eq:gmi1} can be readily obtained for CV states because it allows a phase-space description: i.e., $\mathrm{tr} ( \rho - \sigma )^{2} = \pi^{n} \int d^{2n} \alpha \{ W_{\rho} ( \alpha ) - W_{\sigma}  ( \alpha ) \}^{2}$ for $n$-mode states $\rho$ and $\sigma$, where $W_{\rho} ( \alpha )$ is the Wigner function for the state $\rho$ \cite{Barnett2003}.
As a note, we can also define a fidelity-based correlation, i.e., the quantum Bures mutual information $\mathcal{I}_{\mathrm{B}}^{2} [ \rho_{AB} ] = 2 ( 1 - \sqrt{F [ \rho_{AB}, \rho_{A} \otimes \rho_{B} ]} )$, where $F [ \rho, \sigma ] = ( \mathrm{tr} \sqrt{\sqrt{\rho} \sigma \sqrt{\rho}} )^{2}$ is the quantum fidelity between $\rho$ and $\sigma$ \cite{Uhlmann1976, Jozsa1994}. It is, however, a special case of the second type of quantum R{\'e}nyi mutual information for $\alpha = \frac{1}{2}$ because both of the measures are directly related to the fidelity, i.e., $S_{1/2} [ \rho || \sigma ] = - \ln F [ \rho, \sigma ]$.

\section{Breakdown of Gaussian extremality in quantum-state correlation}
A non-Gaussianity measure for quantum mutual information must satisfy two obvious properties: (P1) The measure is zero if a target state $\rho_{AB}$ has no correlation, i.e., $\rho_{AB} = \rho_{A} \otimes \rho_{B}$, or is Gaussian, i.e., $\rho_{AB} = \sigma_{AB}$ where we denote as $\sigma_{AB}$ the reference Gaussian state. (P2) The measure is invariant under local unitary Gaussian operations, which have no effect on correlation and non-Gaussianity of local and global states. In addition, it is desirable to satisfy two more properties: (P3) The measure is non-negative. (P4) The measure is nonincreasing under local Gaussian channels because they increase neither correlation nor non-Gaussianity of local and global states.

	\begin{figure}[!t]
		\includegraphics[scale=0.55]{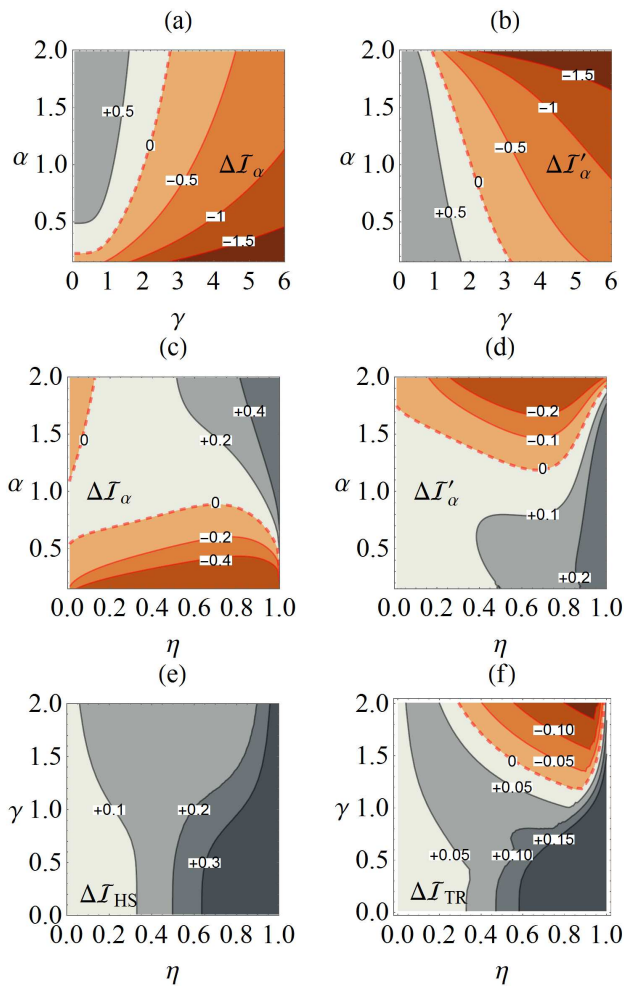}
		\caption{(a), (b) Contour plots of $\Delta \mathcal{I}_{\alpha} = \mathcal{I}_{\alpha} [ \rho_{AB} ] - \mathcal{I}_{\alpha} [ \sigma_{AB} ]$ and $\Delta \mathcal{I}^{\prime}_{\alpha} = \mathcal{I}_{\alpha}^{\prime} ( \rho_{AB} ) - \mathcal{I}_{\alpha}^{\prime} ( \sigma_{AB} )$ for the entangled coherent state $\ket{\Psi}$ against the coherent amplitude $\gamma$ and R{\'e}nyi parameter $\alpha$. Negative regions indicate the breakdown of Gaussian extremality. (c), (d) Contour plots of $\Delta \mathcal{I}_{\alpha}$, $\Delta \mathcal{I}^{\prime}_{\alpha}$ for the entangled coherent state under a symmetric loss channel $\mathcal{L}_{\eta} [ \ketbra{\Psi}{\Psi} ]$ with $\gamma=1$ as functions of the effective transmittance $\eta$ and the R{\'e}nyi parameter $\alpha$.  (e), (f) Contour plots for the loss dynamics of $\Delta \mathcal{I}_{\mathrm{HS}} = \mathcal{I}_{\mathrm{HS}} ( \rho_{AB} ) - \mathcal{I}_{\mathrm{HS}} ( \sigma_{AB} )$ and $\Delta \mathcal{I}_{\mathrm{TR}} = \mathcal{I}_{\mathrm{TR}} ( \rho_{AB} ) - \mathcal{I}_{\mathrm{TR}} ( \sigma_{AB} )$ for the entangled coherent state against the amplitude $\gamma$ and the effective transmittance $\eta$.}
		\label{fig:counterexample1}
	\end{figure}

To quantify the non-Gaussianity of correlation, we may attempt a method of subtracting the correlation of the reference Gaussian state from that of a target state under a given measure, which immediately fulfills the properties (P1) and (P2). If a target state is a product state ($\mathcal{I} [ \rho ] = 0$) or Gaussian, the reference state is also a product state ($\mathcal{I} [ \sigma ] = 0$) or the target state itself ($\rho = \sigma$), respectively. The measures thus satisfy (P1). A Gaussian unitary operation changes the first and second moments of a state according to a linear transformation for position and momentum operators, which is independent of the state \cite{Barnett2003}. If we apply a local Gaussian unitary operation on a target state, its reference state is also affected by the same operation. In addition, the mutual informations [cf. Eqs.~(2),~(3),~(5),~and~(6)] are invariant under local unitary operations. The measures thus meet (P2).

Then, is it possible for such non-Gaussianity measures to meet (P3) and (P4) as well? We give a negative answer by finding counterexamples. 
In Figs. \ref{fig:counterexample1}(a) and \ref{fig:counterexample1}(b), we plot the non-Gaussianity of quantum R{\'e}nyi mutual informations for an entangled coherent state $\ket{\Psi} = \frac{1}{\sqrt{\mathcal{N}}} ( \ket{\gamma}_{A} \ket{\gamma}_{B} - \ket{- \gamma}_{A} \ket{- \gamma}_{B} )$ where $\mathcal{N} = 2 - 2 e^{- 4 | \gamma |^{2}}$ and $\ket{\gamma}$ is the coherent state with amplitude $\gamma$. The quantum mutual information of the entangled coherent state is independent of $\gamma$ since it is essentially a Bell state, i.e., $\ket{\Psi} = \frac{1}{\sqrt{2}} ( \ket{+}_{A} \ket{-}_{B} + \ket{-}_{A} \ket{+}_{B} )$ where $\ket{\pm} = \frac{1}{\sqrt{\mathcal{N_{\pm}}}} ( \ket{\gamma} \pm \ket{- \gamma} )$ with $N_{\pm} = 2 \pm 2 e^{- 2 | \gamma |^{2}}$. Furthermore, the quantum R{\'e}nyi mutual information of a Bell state is $2\ln2$ (2 bits) independent of $\alpha$ (see Appendixes A and B for the quantum mutual information of the entangled coherent state). However, the quantum R{\'e}nyi mutual information of its reference Gaussian state increases with $\gamma$ and can even exceed that of the entangled coherent state (refer to Appendixes C and D for the quantum mutual information of the reference Gaussian state).

In Figs. \ref{fig:counterexample1}(c)-\ref{fig:counterexample1}(f), we also examine the dynamics of the non-Gaussianity measures under a symmetric loss channel $\mathcal{L}_{\eta}$. A loss channel can be modeled by the mixing of an input state and vacuum at a beam-splitter with transmittance $\eta$. Here we focus on the case that each mode passes through the same loss channel. The contour plots manifest that the three measures $\Delta \mathcal{I}_{\alpha}$, $\Delta \mathcal{I}_{\alpha}^{\prime}$ and $\Delta \mathcal{I}_{\mathrm{TR}}$ are neither non-negative nor monotonic under local Gaussian channels.

	\begin{figure}[!t]
		\includegraphics[scale=0.55]{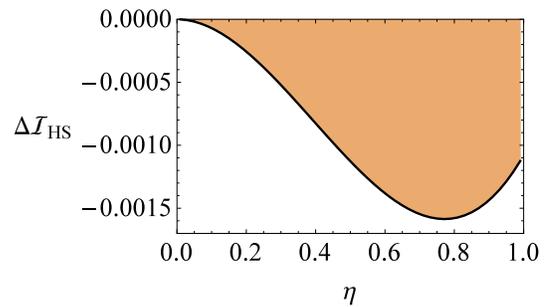}
		\caption{Loss dynamics of $\Delta \mathcal{I}_{\mathrm{HS}} = \mathcal{I}_{\mathrm{HS}} ( \rho_{AB} ) - \mathcal{I}_{\mathrm{HS}} ( \sigma_{AB} )$ for a photon number entangled state $\ket{\psi} = \sum_{k=0}^{2} c_{k} \ket{k}_{A} \ket{k}_{B}$ with $c_{0} = 0.986$, $c_{1} = 0.162$, and $c_{2} = (1-c_{0}^{2}-c_{1}^{2})^{1/2}$ as a function of the effective transmittance $\eta$.}
		\label{fig:counterexample2}
	\end{figure}

In Fig. \ref{fig:counterexample1}(e), on the other hand, the non-Gaussianity of the Hilbert-Schmidt mutual information $\Delta \mathcal{I}_{\mathrm{HS}}$ seems well behaved, unlike other measures, with monotonicity under local Gaussian channels. However, we can also observe the breakdown of Gaussian extremality of $\mathcal{I}_{\mathrm{HS}}$ by using a different state, i.e., a photon number entangled state in the form $\sum_{k} c_{k} \ket{k}_{A} \ket{k}_{B}$ \cite{Lee2012}. In Fig.~\ref{fig:counterexample2}, we plot the loss dynamics of $\Delta \mathcal{I}_{\mathrm{HS}}$ for $\sum_{k=0}^{2} c_{k} \ket{k}_{A} \ket{k}_{B}$ with $c_{0} = 0.986$, $c_{1} = 0.162$, and $c_{2} = (1-c_{0}^{2}-c_{1}^{2})^{1/2} \simeq 0.039$. It clearly shows that $\Delta \mathcal{I}_{\mathrm{HS}}$ also fails to meet (P3) and (P4).


	\begin{figure}[!t]
		\includegraphics[scale=0.55]{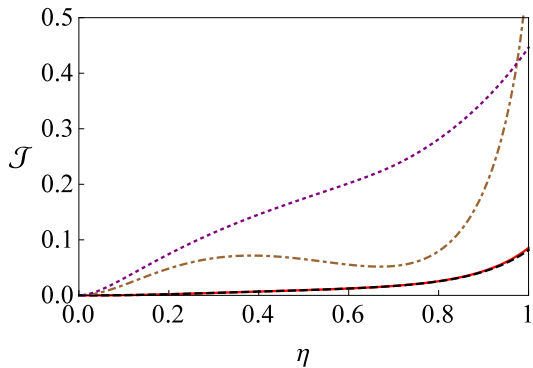}
		\caption{$\mathcal{J}_{D}$ (trace distance, purple dotted curve), $\mathcal{J}_{\mathrm{LB1}}$ (black dashed curve) and $\mathcal{J}_{\mathrm{LB2}}$ (red solid curve) for the entangled coherent state with amplitude $\gamma=1$ under the symmetric loss channel $\mathcal{L}_{\eta}$ plotted against the effective transmittance $\eta$. The difference between the two measures $\mathcal{J}_{\mathrm{LB1}}$ and $\mathcal{J}_{\mathrm{LB2}}$ is negligible. For comparison, we also plot $\Delta \mathcal{I}_{1}$ (brown dot-dashed curve)  which clearly shows a nonmonotonic behavior.}
		\label{fig:SNG}
	\end{figure}
		
\section{Non-Gaussianity measure for quantum mutual information} \label{sec:NEW}
In this section, we propose a method to quantify non-Gaussianity of correlation satisfying all necessary properties introduced in Sec. III.
Inspired from Eqs.~\eqref{eq:gmi1} and~\eqref{eq:gmi}, it makes sense to look into the difference between a global state and the product of local states, i.e., $\rho_{AB} - \rho_{A} \otimes \rho_{B}$, in order to obtain information on correlation. It may be plausible to define the non-Gaussianity of correlation as the distance between the correlation parts of the target state and its reference state, i.e., $D [ \rho_{AB} - \rho_{A} \otimes \rho_{B}, \sigma_{AB} - \sigma_{A} \otimes \sigma_{B} ]$, for any legitimate measure of distance $D$. However, we note that the correlation parts themselves do not represent physical states because $\mathrm{tr} [ \rho_{AB} - \rho_{A} \otimes \rho_{B} ] = \mathrm{tr} [ \sigma_{AB} - \sigma_{A} \otimes \sigma_{B} ] = 0$. We can overcome this issue by defining the distance in terms of two physical states by using the property $D [ \rho_{AB} - \rho_{A} \otimes \rho_{B}, \sigma_{AB} - \sigma_{A} \otimes \sigma_{B} ] = D [ \rho_{AB} + \sigma_{A} \otimes \sigma_{B}, \sigma_{AB} + \rho_{A} \otimes \rho_{B} ]$. We thus introduce two relevantly averaged states, i.e., 
\begin{eqnarray}
\widetilde{\rho}_{AB} &=& \frac{1}{2} ( \rho_{AB} + \sigma_{A} \otimes \sigma_{B} )\nonumber\\
\widetilde{\sigma}_{AB} &=& \frac{1}{2} ( \sigma_{AB} + \rho_{A} \otimes \rho_{B} ), 
\end{eqnarray}
and define the measure of non-Gaussian correlation as
	\begin{equation} \label{eq:N2}
		\mathcal{J}_{D} [ \rho_{AB} ] \equiv D [ \widetilde{\rho}_{AB}, \widetilde{\sigma}_{AB} ].
	\end{equation}
It may be regarded as a smoothed non-Gaussianity measure for $D [ \rho_{AB} - \rho_{A} \otimes \rho_{B}, \sigma_{AB} - \sigma_{A} \otimes \sigma_{B} ]$ obtained by jointly mixing $\rho_{A} \otimes \rho_{B} + \sigma_{A} \otimes \sigma_{B}$ into each correlation part.

Equation \eqref{eq:N2} generates a non-Gaussianity measure satisfying all the properties introduced in Sec. III if the distance measure $D$ is invariant under unitary operations and nonincreasing under quantum channels, e.g., trace distance. The properties (P1) and (P2) are readily shown to satisfy in the same way we used in the previous section. The property (P3) is also straightforwardly satisfied because a distance measure is non-negative by definition. We thus need to focus on property (P4) only. Every Gaussian channel can be described as an interaction with a Gaussian environment, i.e., $\mathcal{M} [ \rho ] \equiv \mathrm{tr}_{E} [ \hat{U}_{G} \rho \otimes \sigma_{E} \hat{U}_{G}^{\dag} ]$ \cite{Cerf2007}. Combining it with the argument for (P2) in the previous section, we find that, a target state and its reference state evolve under the same Gaussian channel $\mathcal{M}$. When a local Gaussian channel, $\mathcal{M} = \mathcal{M}_{A} \otimes \mathcal{M}_{B}$, is applied, Eq.~\eqref{eq:N2} becomes $D [ \mathcal{M} [ \widetilde{\rho}_{AB} ], \mathcal{M} [ \widetilde{\sigma}_{AB} ] ]$. Therefore, if the distance measure $D$ is nonincreasing under quantum channels, (P4) is satisfied.

Using the trace-distance as a quantifier of correlation, we plot in Fig. 4 its behavior as a function of $\eta$ (transmittance) of the lossy Gaussian channel for the case of an entangled coherent state (purple dotted curve). It shows a monotonically decreasing behavior with loss, in contrast to the nonmonotonic behavior of $\Delta \mathcal{I}_{1}$ (brown dot-dashed curve). 

We may define the non-Gaussian correlation measures based on other quantities like R{\'e}nyi entropy or fidelity. 
The quantum R{\'e}nyi relative entropy with $\alpha \geq \frac{1}{2}$ satisfies desirable conditions like the nonincreasing behavior under quantum channels \cite{Frank2013}. However, it may be challenging to directly compute these measures as expanding the mixtures of non-Gaussian and Gaussian states in the eigen-decomposition can generally be nontrivial. Coping with the difficulty, we focus on the fidelity-based measures providing readily computable lower bounds. 
In Refs.~\cite{Mendonca2008, Miszczak2009}, an alternative fidelity measure is proposed as
	\begin{equation}
		G [ \rho, \sigma ] = \mathrm{tr} [ \rho \sigma ] + \sqrt{1 - \mathrm{tr} [ \rho^{2} ]} \sqrt{1 - \mathrm{tr} [ \sigma^{2} ]},
	\end{equation}
which is an upper bound of the conventional fidelity measure, i.e., $F \leq G$. Interestingly, this superfidelity measure is computable for an arbitrary pair of quantum states because the measure allows a phase-space description: $\mathrm{tr} [ \rho \sigma ] = \pi^{n} \int d^{2n} \alpha W_{\rho} ( \alpha ) W_{\sigma} ( \alpha )$. Contrary to the fidelity, it is unnecessary for the superfidelity to solve an eigenvalue problem on an infinite-dimensional Hilbert space.

We now define our computable non-Gaussianity measure as 
\begin{eqnarray}
\mathcal{J}_{\mathrm{LB1}} [ \rho_{AB} ] = - \ln G [ \widetilde{\rho}_{AB}, \widetilde{\sigma}_{AB} ]
\end{eqnarray}
in relation to $S_{1/2} [ \rho || \sigma ] = - \ln F [ \rho, \sigma ]$. In addition, we propose another computable measure based on quantum Hilbert-Schmidt distance as
\begin{eqnarray}
\mathcal{J}_{\mathrm{LB2}} [ \rho_{AB} ] = - \ln ( 1 - \frac{1}{2} D_{\mathrm{HS}}^{2} [ \widetilde{\rho}_{AB}, \widetilde{\sigma}_{AB} ] ),
\end{eqnarray}
satisfying $\mathcal{J}_{\mathrm{LB1}} [ \rho_{AB} ] \geq \mathcal{J}_{\mathrm{LB2}} [ \rho_{AB} ]$. The order relation between two measures can be seen from
	\begin{align}
		G [ \rho_{1}, \rho_{2} ] & \leq \mathrm{tr} [ \rho_{1} \rho_{2} ] + \frac{( 1 - \mathrm{tr} [ \rho^{2} ] ) + ( 1 - \mathrm{tr} [ \sigma^{2} ] )}{2} \nonumber \\
		& = 1 - \frac{1}{2} \mathrm{tr} ( \rho_{1} - \rho_{2} )^{2} \nonumber \\
		& = 1 - \frac{1}{2} D_{\mathrm{HS}}^{2} [ \rho_{1}, \rho_{2} ].
	\end{align}

	\begin{figure}[t]
		\includegraphics[scale=0.55]{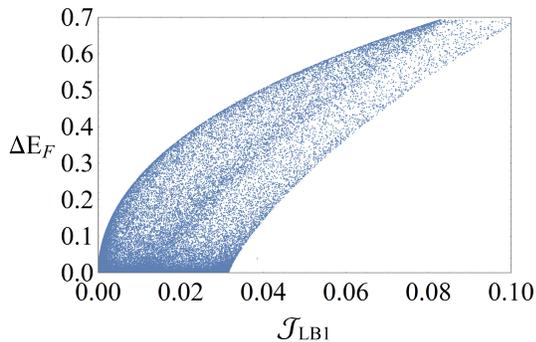}
		\caption{$\Delta E_{F} = E_{F} [ \rho_{AB} ] - E_{F} [ \sigma_{AB} ]$ for the entangled coherent state under a symmetric loss channel versus our non-Gaussianity measure $\mathcal{J}_{\mathrm{LB1}}$. We randomly sampled $10^{5}$ states for this parametric plot and observe a positive relation between two measures.}
		\label{fig:EF}
	\end{figure}

It also implies that $G [ \rho_{1}, \rho_{2} ]$ is a strict fidelity measure taking the value of unity {\it iff} the two states are identical and less than unity otherwise, and $\mathcal{J}_{\mathrm{LB1}}$ and $\mathcal{J}_{\mathrm{LB2}}$ meet (P3). However, the lower bounds may fail to satisfy (P4) because the superfidelity and quantum Hilbert-Schmidt distance can be increased under a quantum channel \cite{Mendonca2008}. Therefore, strictly speaking, the quantities in Eqs. (10) and (11) can be used only as a lower bound to (under)estimate the true measure based on a regular fidelity. Of course, we must take care in interpreting these lower bounds; for instance, we cannot determine which state possesses a stronger non-Gaussian correlation by comparing only the lower bounds among different states. 

Two measures $\mathcal{J}_{\mathrm{LB1}}$ and $\mathcal{J}_{\mathrm{LB2}}$ do not show an appreciable difference in many cases because the gap between the arithmetic and geometrical means of impurities, i.e., $\frac{\widetilde{\mu}_{1} + \widetilde{\mu}_{2}}{2}$ and $\sqrt{\widetilde{\mu}_{1} \widetilde{\mu}_{2}}$ where $\widetilde{\mu}_{i} = 1 - \mathrm{tr} \rho_{i}^{2}$ with $i \in \{ 1, 2 \}$, is small for a wide range of parameters. Note that the two averaged states $\widetilde{\rho}_{AB}$ and $\widetilde{\sigma}_{AB}$ can be pure {\it iff} $\rho_{A}$ and $\rho_{B}$ are pure Gaussian states.

We plot $\mathcal{J}_{\mathrm{LB1}}$ (black dashed curve) and $\mathcal{J}_{\mathrm{LB2}}$ (red solid curve) for the entangled coherent state in Fig.~\ref{fig:SNG}. We see that the lower bounds in Eqs. (10) and (11) show a monotonic behavior under a local Gaussian channel using our approach.

	\begin{figure}[t]
		\includegraphics[scale=0.55]{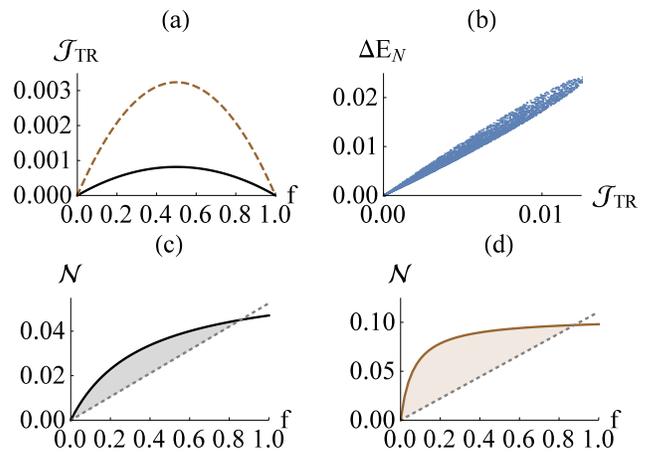}
		\caption{(a) $\mathcal{J}_{\mathrm{TR}}$ for CV Werner state in Eq.~\eqref{eq:CVWS} against the fraction $f$. Black solid and brown dashed curves represent the cases for $r=0.05$ and $r=0.1$, respectively. (b) $\Delta E_{N} = E_{N} [ \rho_{AB} ] - E_{N} [ \sigma_{AB} ]$ versus our non-Gaussianity measure $\mathcal{J}_{\mathrm{TR}}$ for CV Werner states. We have randomly sampled $10^{4}$ states with $0 \leq f \leq 1$ and $0 \leq r \leq 0.2$ for this parametric plot. (c), (d) entanglement negativity $\mathcal{N}$ for the original CV Werner state (dashed) and an output state from a Gaussian distillation protocol (solid), respectively. [(c) $r$=0.05 and (d) $r$=0.1] Shaded regions represent the degree of distillable entanglement with respect to the fraction $f$.}
		\label{fig:CVWS}
	\end{figure}

Note that $\mathcal{J}_{\mathrm{LB2}}$ is particularly useful for two extremal cases: (i) The reference Gaussian state is a product state, i.e., $\sigma_{AB} = \sigma_{A} \otimes \sigma_{B}$. (ii) The target state is a non-Gaussian correlated state with local Gaussian states. For these cases, we can further reduce the computational efforts because only two states are involved, i.e., $\mathcal{J}_{\mathrm{LB2}} [ \rho_{AB} ] = - \ln ( 1 - \frac{1}{8} D_{\mathrm{HS}}^{2} [ \rho_{AB}, \rho_{A} \otimes \rho_{B} ] )$ and $\mathcal{J}_{\mathrm{LB2}} [ \rho_{AB} ] = - \ln ( 1 - \frac{1}{8} D_{\mathrm{HS}}^{2} [ \rho_{AB}, \sigma_{AB} ] )$ for (i) and (ii), respectively. 
We provide explicit examples for the extremal cases. (i) A photon number entangled state in the form $\ket{\psi} = \sum_{k} c_{k} \ket{n_{k}}_{A} \ket{n_{k}}_{B}$ satisfying $n_{k + 1} - n_{k} \geq 2$ for every $k$, e.g., $\ket{\psi} = \sqrt{x} \ket{0}_{A} \ket{0}_{B} + \sqrt{1-x} \ket{2}_{A} \ket{2}_{B}$, has no Gaussian correlation because its covariance matrix is $\Gamma = ( \bar{n} + \frac{1}{2} ) \mathbb{I}_{4}$ with $\bar{n} = \sum_{k} n_{k} | c_{k} |^{2}$. (ii) A photon number correlated state in the form $\rho = \sum_{k = 0}^{\infty} \frac{\bar{n}^{k}}{(\bar{n} + 1)^{k+1}} \ketbra{k}{k}_{A} \otimes \ketbra{k}{k}_{B}$ is locally Gaussian but globally non-Gaussian. We identify the global state as non-Gaussian with its covariance matrix equivalent to that of a uncorrelated thermal state, i.e., $\Gamma = ( \bar{n} + \frac{1}{2} ) \mathbb{I}_{4}$.



In Fig.~\ref{fig:EF}, we consider the difference in the entanglement of formation \cite{Wooters1998} between the target state and the reference Gaussian state, $\Delta E_{F} = E_{F} [ \rho_{AB} ] - E_{F} [ \sigma_{AB} ]$, and compare it with our non-Gaussianity measure for the entangled coherent state under a symmetric loss channel $\mathcal{L}_{\eta}$. From the Gaussian entanglement criterion \cite{Duan2000, Simon2000}, we observe that the reference Gaussian state has no entanglement ($E_{F} [\sigma_{AB}]=0$), which allows us to focus on the entanglement of formation for the target state with a two-qubit structure \cite{Wooters1998}. Although there does not exist a one-to-one correspondence in defining those two measures, the amount of entanglement and our non-Gaussianity measure show a positive relation with a bounded interval. Of course, such an interesting feature disappears if we look into non-Gaussianity of total state, e.g., $S_{1} ( \sigma_{AB} ) - S_{1} ( \rho_{AB} )$, instead of our measures.

Let us now consider another example, i.e., CV Werner states \cite{Mista2002}; that is, a mixture of a vacuum state and a two-mode squeezed state,
	\begin{equation} \label{eq:CVWS}
		\rho_{AB} = (1-f) \ketbra{0}{0}_{A} \otimes \ketbra{0}{0}_{B} + f \ketbra{\phi}{\phi},
	\end{equation}
where $\ket{\phi} = \frac{1}{\cosh r} \sum_{k=0}^{\infty} \tanh^{k} r \ket{k}_{A} \ket{k}_{B}$ is a two-mode squeezed vacuum with squeezing strength $r$. Here we focus on experimentally feasible cases with weak squeezing $r \leq 0.2$ \cite{Kurochkin2014}. We investigate the difference in the entanglement negativity \cite{Vidal2002, Plenio2005} between the given state and the reference Gaussian state, $\Delta E_{N} = E_{N} [ \rho_{AB} ] - E_{N} [ \sigma_{AB} ]$, in comparison with our non-Gaussianity measure $\mathcal{J}_{\mathrm{TR}}$ in Fig.~\ref{fig:CVWS}, which again shows a strong correlation between two measures.

We may further seek to identify the role of non-Gaussian correlation in other quantum tasks, e.g., distillable entanglement by Gaussian operations. To this aim, we introduce here a Gaussian distillation protocol by using beam splitters and homodyne measurements as follows: (1) We prepare a non-Gaussian correlated state, e.g., $\rho_{AB} = (1-f) \ketbra{0}{0}_{A} \otimes \ketbra{0}{0}_{B} + f \ketbra{\phi}{\phi}$. (2) Each mode passes through a beam splitter $\hat{\mathcal{B}}$ with transmittance $\eta$, i.e., $\widetilde{\rho}_{ABCD} = \hat{\mathcal{B}}_{AC} \hat{\mathcal{B}}_{BD} ( \rho_{AB} \otimes \ketbra{0}{0}_{C} \otimes \ketbra{0}{0}_{D} ) \hat{\mathcal{B}}_{BD}^{\dag} \hat{\mathcal{B}}_{AC}^{\dag}$. (3) We measure the quadratures of the ancillary modes C and D, and postselect the outcomes of homodyne measurements. It yields
	\begin{equation} \label{eq:DS}
		\rho_{AB}^{\prime} = \frac{\bra{x_{c}} \bra{x_{d}} \widetilde{\rho}_{ABCD} \ket{x_{c}} \ket{x_{d}}}{\mathrm{tr} [ \bra{x_{c}} \bra{x_{d}} \widetilde{\rho}_{ABCD} \ket{x_{c}} \ket{x_{d}} ]},
	\end{equation}
where $\ket{x_{c}}$ and $\ket{x_{d}}$ represent quadrature eigenstates \cite{Barnett2003} for the ancillary modes C and D, respectively. 

In Figs.~\ref{fig:CVWS}(c) and \ref{fig:CVWS}(d), we plot $E_{N}$ for the CV Werner state (dashed) and the output distilled state $\rho_{AB}^{\prime}$ (solid), respectively, for the case of $\eta = 0.9$, $x_{c} = x_{d} = 0.8$. The shaded region represents the degree of distillable entanglement, which conincides overall with our measure of non-Gaussian correlation $\mathcal{J}_{\mathrm{TR}}$ in Fig.~\ref{fig:CVWS}(a). While our current approach does not provide a rigorous basis to make an operational interpretation of our non-Gaussian correlation measure in relation to distillable entanglement, such a relation could be justified under certain conditions, which will be an interesting issue for further investigation.

\section{Conclusion}
We have investigated how to characterize the non-Gaussianity of quantum-state correlation by using various measures such as the relative entropy and geometric distances. We have found that Gaussian extremality holds for none of these measures in our consideration. In other words, there exists a non-Gaussian correlated state that has equal or less quantum mutual information compared with its reference state, i.e., the Gaussian state having the same covariance matrix as the non-Gaussian state. The same issue on Gaussian extremality may be raised toward other quantum correlation measures, such as quantum discord \cite{Ollivier2001}, which would be a topic of further investigation.	

To come up with a measure of non-Gaussian correlation satisfying all desirable properties, we have established a method to characterize the non-Gaussianity contained in correlation part and proposed the distance-based and the fidelity-based measures. For the latter case, we also provided two readily computable lower bounds. It is an issue of crucial importance to characterize and quantify non-Gaussianity for CV quantum informatics, particularly the non-Gaussian quantum correlation. It may be interesting to extend our consideration to the realms of strictly nonclassical correlation and genuine multipartite correlations \cite{Giorgi2011} and investigate their roles for the emergence of advantages in using non-Gaussian resources for CV information processing.

\acknowledgements
This work is supported by the NPRP Grant No. 8-751-1-157 from the Qatar National Research Fund and the R\&D Convergence Program of NST (National Research Council of Science and Technology) of Republic of Korea (Grant No. CAP-18-08-KRISS).

\appendix
\section{Entangled Coherent State}
We first introduce two basis states as
	\begin{equation}
		\ket{\pm} = \frac{1}{\sqrt{N_{\pm}(\gamma)}} ( \ket{\gamma} \pm \ket{- \gamma} ),
	\end{equation}
where $\ket{\gamma}$ is a coherent state with a real amplitude $\gamma$ and the normalization factors given by $N_{\pm} (\gamma) = 2 \pm 2 \exp ( - 2 \gamma^{2} )$. 
The basis states are orthogonal, $\braket{+}{-} = 0$, and can be transformed to each other by a single photon subtraction as
	\begin{equation} \label{eq:transformation}
		\hat{a} \ket{\pm} = \sqrt{\frac{N_{\mp}}{N_{\pm}}} \gamma \ket{\mp}.
	\end{equation}

We then consider an entangled coherent state, $\ket{\Psi} = \frac{1}{\sqrt{2}} ( \ket{+}_{A} \ket{-}_{B} + \ket{-}_{A} \ket{+}_{B} )$, which is one of Bell states. 
The reduced density matrix of a local mode $i \in \{ A, B \}$ for state $\ket{\Psi}$ becomes a maximally mixed state, $\rho_{i} = \frac{1}{2} ( \ketbra{+}{+} + \ketbra{-}{-} )$.

For a pure state of the form $\ket{\psi} = \sum_{k} c_{k} \ket{k}_{A} \ket{k}_{B}$ with $\sum_{k} |c_{k}|^{2} = 1$, the quantum mutual information is given by
	\begin{align}
		\mathcal{I}_{\alpha} [ \ketbra{\psi}{\psi} ] & = \frac{2}{1 - \alpha} \ln \bigg( \sum_{k} |c_{k}|^{2 \alpha} \bigg), \nonumber \\
		\mathcal{I}_{\alpha}^{\prime} [ \ketbra{\psi}{\psi} ] & = \frac{\alpha}{\alpha - 1} \ln \bigg( \sum_{k} |c_{k}|^{\frac{4-2\alpha}{\alpha}} \bigg), \nonumber \\
		\mathcal{I}_{\mathrm{HS}} [ \ketbra{\psi}{\psi} ] & = \sqrt{1 + \bigg( \sum_{k} |c_{k}|^{4} \bigg)^{2} - 2 \bigg( \sum_{k} |c_{k}|^{6} \bigg)},
	\end{align}
which yield $\mathcal{I}_{\alpha} [ \ketbra{\Psi}{\Psi} ] = \mathcal{I}_{\alpha}^{\prime} [ \ketbra{\Psi}{\Psi} ] = 2 \ln 2$ for the case of entangled coherent states.

Under a symmetric loss with effective transmittance $\eta$, the entangled coherent state $\ket{\Psi}$ evolves as
	\begin{align}
		\mathcal{L}_{\eta} [ \ketbra{\Psi}{\Psi} ] = & \frac{N_{+} ( \sqrt{2-2\eta} \gamma ) N_{-} ( \sqrt{2\eta} \gamma )}{4 N_{-} ( \sqrt{2} \gamma )} \ketbra{\Psi^{\prime}}{\Psi^{\prime}} \nonumber \\
		+ & \frac{N_{-} ( \sqrt{2-2\eta} \gamma ) N_{+} ( \sqrt{2\eta} \gamma )}{4 N_{-} ( \sqrt{2} \gamma )} \ketbra{\Xi^{\prime}}{\Xi^{\prime}},
	\end{align}
where
	\begin{align}
		\ket{\Psi^{\prime}} & = \frac{1}{\sqrt{2}} ( \ket{+^{\prime}}_{A} \ket{-^{\prime}}_{B} + \ket{-^{\prime}}_{A} \ket{+^{\prime}}_{B} ), \nonumber \\
		\ket{\Xi^{\prime}} & = \frac{N_{+} ( \sqrt{\eta} \gamma )}{2 \sqrt{N_{+} ( \sqrt{2 \eta} \gamma )}}  \ket{+^{\prime}}_{A} \ket{+^{\prime}}_{B} \nonumber \\
		& + \frac{N_{-} ( \sqrt{\eta} \gamma )}{2 \sqrt{N_{+} ( \sqrt{2 \eta} \gamma )}}  \ket{-^{\prime}}_{A} \ket{-^{\prime}}_{B},
	\end{align}
with
	\begin{equation}
		\ket{\pm^{\prime}} = \frac{1}{\sqrt{N_{\pm} ( \sqrt{\eta} \gamma )}} ( \ket{\sqrt{\eta} \gamma} \pm \ket{- \sqrt{\eta} \gamma} ).
	\end{equation}

\section{Mutual information for two qubit X states}
For a two-qubit X state in the form,
	\begin{equation}
		\rho =
		\begin{pmatrix}
			a & & & v \\
			& b & u & \\
			& u^{*} & c & \\
			v^{*} & & & d
		\end{pmatrix}
	\end{equation}
with basis states $\{ \ket{+} \ket{+}, \ket{+} \ket{-}, \ket{-} \ket{+}, \ket{-} \ket{-} \}$, the first type of quantum R{\'e}nyi mutual information is given by
	\begin{align}
		\mathcal{I}_{\alpha} [ \rho ] & = \frac{1}{1-\alpha} \ln \{ (a+b)^{\alpha} + (c+d)^{\alpha} \} \nonumber \\
		& + \frac{1}{1-\alpha} \ln \{ (a+c)^{\alpha} + (b+d)^{\alpha} \} \nonumber \\
		& - \frac{1}{1-\alpha} \ln \{ \sum_{i=1}^{4} \lambda_{i}^{\alpha} \},
	\end{align}
where
	\begin{align}
		\lambda_{1,2} & = \frac{1}{2} ( a+d \pm \sqrt{(a-d)^{2}+4|v|^{2}} ), \nonumber \\
		\lambda_{3,4} & = \frac{1}{2} ( b+c \pm \sqrt{(b-c)^{2}+4|u|^{2}} ).
	\end{align}
On the other hand, the second type of quantum R{\'e}nyi mutual information is given by
	\begin{align}
		\mathcal{I}_{\alpha}^{\prime} [ \rho ] & = \frac{1}{\alpha-1} \ln \{ \sum_{i=1}^{4} (\lambda_{i}^{\prime})^{\alpha} \},
	\end{align}
where
	\begin{align}
		\lambda_{1,2}^{\prime} & = \frac{1}{2} ( a^{\prime} + d^{\prime} \pm \sqrt{(a^{\prime}-d^{\prime})^{2}+4|v^{\prime}|^{2}} ), \nonumber \\
		\lambda_{3,4}^{\prime} & = \frac{1}{2} ( b^{\prime} + c^{\prime} \pm \sqrt{(b^{\prime}-c^{\prime})^{2}+4|u^{\prime}|^{2}} ),
	\end{align}
with
	\begin{align}
		a^{\prime} & = a (a+b)^{\frac{1-\alpha}{\alpha}} (a+c)^{\frac{1-\alpha}{\alpha}}, \nonumber \\
		b^{\prime} & = b (a+b)^{\frac{1-\alpha}{\alpha}} (b+d)^{\frac{1-\alpha}{\alpha}}, \nonumber \\
		c^{\prime} & = c (c+d)^{\frac{1-\alpha}{\alpha}} (a+c)^{\frac{1-\alpha}{\alpha}}, \nonumber \\
		d^{\prime} & = d (c+d)^{\frac{1-\alpha}{\alpha}} (b+d)^{\frac{1-\alpha}{\alpha}}, \nonumber \\
		u^{\prime} & = u \{(a+b)(c+d)(a+c)(b+d)\}^{\frac{1-\alpha}{2\alpha}}, \nonumber \\
		v^{\prime} & = v \{(a+b)(c+d)(a+c)(b+d)\}^{\frac{1-\alpha}{2\alpha}}.
	\end{align}
Finally, the quantum Hilbert-Schmidt mutual information for the two-qubit X state is given by
	\begin{align}
		\mathcal{I}_{\mathrm{HS}}^{2} [ \rho ] = & \{a-(a+b)(a+c)\}^{2} + \{b-(a+b)(b+d)\}^{2} \nonumber \\
		+ & \{c-(c+d)(a+c)\}^{2} + \{d-(c+d)(b+d)\}^{2} \nonumber \\
		+ & 2 ( |u|^{2} + |v|^{2} ).
	\end{align}

\section{Mutual Information for Gaussian States}
An $n$-mode Gaussian state is characterized by the first-order moments, i.e., $\expect{\hat{Q}} = ( \expect{\hat{q}_{1}}, \expect{\hat{p}_{1}}, ... \expect{\hat{q}_{n}}, \expect{\hat{p}_{n}} )$, and its covariance matrix $\Gamma$ whose elements are defined as
	\begin{equation}
		\Gamma_{ij} = \frac{1}{2} \expect{\hat{Q}_{i} \hat{Q}_{j} + \hat{Q}_{j} \hat{Q}_{i}} - \expect{\hat{Q}_{i}} \expect{\hat{Q}_{j}},
	\end{equation}
where $\hat{q}_{j} = \frac{\hat{a}_{j}+\hat{a}_{j}^{\dag}}{\sqrt{2}}$ and $\hat{p}_{j} = \frac{\hat{a}_{j}-\hat{a}_{j}^{\dag}}{\sqrt{2}i}$ are the position and momentum operators for the $j$th mode, respectively.

For a Gaussian state $\sigma$, there always exists a Gaussian unitary operator $\hat{S}$ that transforms the state $\sigma$ into a product of thermal state \cite{Adesso2014}: $\hat{S} \sigma \hat{S}^\dag = \bigotimes_{j=1}^{n} \sigma ( \lambda_{j} - \frac{1}{2} )$ where $\sigma_{\mathrm{th}} ( \bar{n} ) = \sum_{k = 0}^{\infty} \frac{\bar{n}^{k}}{( \bar{n} + 1 )^{k + 1}} \ketbra{k}{k}$ represents a thermal state with the mean photon number $\bar{n}$,  and $\lambda_{j}$ is the $j$th symplectic eigenvalue of the covariance matrix $\Gamma$, i.e., the positive eigenvalues of the matrix $i \Omega \Gamma$ with $\Omega = \bigoplus_{j=1}^{n} \omega$ and $\omega = \begin{pmatrix} 0 & 1 \\ -1 & 0 \end{pmatrix}$.
Using the same transformation, we obtain
	\begin{equation} \label{eq:GP} 
		\frac{\hat{S} \sigma^{\alpha} \hat{S}^{\dag}}{\mathrm{tr} [ \sigma^{\alpha} ]} = \bigoplus_{j=1}^{n} \sigma_{\mathrm{th}} ( \zeta_{j} - \frac{1}{2} ),
	\end{equation}
where
	\begin{equation} 
		\zeta_{j} = \frac{1}{2} \frac{( \lambda_{j} + \frac{1}{2} )^{\alpha} + ( \lambda_{j} - \frac{1}{2} )^{\alpha}}{( \lambda_{j} + \frac{1}{2} )^{\alpha} - ( \lambda_{j} - \frac{1}{2} )^{\alpha}},
	\end{equation}
and $\mathrm{tr} [ \sigma^{\alpha} ] = \prod_{j=1}^{n} g ( \lambda_{j}, \alpha )$ with
	\begin{equation} 
		g ( x, \alpha ) = \frac{1}{( x + \frac{1}{2} )^{\alpha} - ( x - \frac{1}{2} )^{\alpha}}.
	\end{equation}

From now on, without loss of generality, let us deal only with the standard form of the covariance matrix for the calculation of quantum mutual informations, i.e.,
	\begin{equation}
		\Gamma = \begin{pmatrix} a & & c & \\ & a & & d \\ c & & b & \\ & d & & b \end{pmatrix}.
	\end{equation}
Note that there always exists a local Gaussian unitary operator that transforms the covariance matrix of a two-mode state into its standard form, and the unitary operation has no effect on the mutual information measures. 
The first type of R{\'e}nyi mutual information $\mathcal{I}_{\alpha} [ \sigma_{AB} ]$ is obtained as
	\begin{equation}
		\mathcal{I}_{\alpha} [ \sigma_{AB} ] = \frac{1}{1-\alpha} \ln \frac{g ( a, \alpha ) g ( b, \alpha )}{g ( \lambda_{1}, \alpha ) g ( \lambda_{2}, \alpha )},
	\end{equation}
where $\lambda_{1,2} = [ \ell\pm (\ell^{2}+m)^{1/2}]^{1/2}$ with $\ell=\frac{1}{2}(a^{2}+b^{2}+2cd)$ and $m=(ab-c^{2})(ab-d^{2})$.

Next, we introduce a composition rule for two Gaussian states having the same means \cite{Marian2012} as
	\begin{equation} \label{eq:CPR} 
		\sigma_{1} \sigma_{2} = \frac{1}{\sqrt{\det ( \Gamma_{1} + \Gamma_{2} )}} \widetilde{\sigma},
	\end{equation}
where $\widetilde{\sigma}$ is a Gaussian state with the same mean and its covrainace matrix is obtained by
	\begin{align} 
		& h ( \Gamma_{1}, \Gamma_{2} ) \nonumber \\
		& =  - \frac{i}{2} \Omega + ( \Gamma_{2} + \frac{i}{2} \Omega ) ( \Gamma_{1} + \Gamma_{2} )^{-1} ( \Gamma_{1} + \frac{i}{2} \Omega ).
	\end{align}

Here we derive a computable expression of $\mathcal{I}_{\alpha}^{\prime} [ \sigma_{AB} ]$ for a Gaussian state as follows: Using the transformation in Eq.~\eqref{eq:GP}, we first have 
	\begin{align}
		& \mathcal{I}_{\alpha}^{\prime} [ \sigma_{AB} ] \nonumber \\
		& = \frac{1}{\alpha-1} \ln \mathrm{tr} [ \{ (\sigma_{A} \otimes \sigma_{B})^{\frac{1-\alpha}{2\alpha}} \sigma_{AB} (\sigma_{A} \otimes \sigma_{B})^{\frac{1-\alpha}{2\alpha}} \}^{\alpha} ] \nonumber \\
		& = \frac{2\alpha}{\alpha-1} \ln [ g ( a, \frac{1-\alpha}{2\alpha} ) g ( b, \frac{1-\alpha}{2\alpha} ) ] \nonumber \\
		& + \frac{1}{\alpha-1} \ln \mathrm{tr} [ \{ ( \sigma_{A}^{\prime} \otimes \sigma_{B}^{\prime} ) \sigma_{AB} ( \sigma_{A}^{\prime} \otimes \sigma_{B}^{\prime} ) \}^{\alpha} ],
	\end{align}
where $\sigma^{\prime}_{j} = \frac{\sigma_{j}^{\alpha}}{\mathrm{tr} [ \sigma_{j}^{\alpha} ]}$ with $j \in \{ A, B \}$. Employing the composition rule in Eq.~\eqref{eq:CPR} twice and Eq.~\eqref{eq:GP} again, we obtain
	\begin{align}
		& \mathcal{I}_{\alpha}^{\prime} [ \sigma_{AB} ] \nonumber \\
		& = \frac{2\alpha}{\alpha-1} \ln [ g ( a, \frac{1-\alpha}{2\alpha} ) g ( b, \frac{1-\alpha}{2\alpha} ) ] \nonumber \\
		& - \frac{\alpha}{2(\alpha-1)} \ln [ \det ( \Gamma^{\prime} + \Gamma ) \det \{ h ( \Gamma^{\prime}, \Gamma ) + \Gamma^{\prime} \} ] \nonumber \\
		& + \frac{1}{\alpha - 1} \ln [ g ( \widetilde{\lambda}_{1}, \alpha ) g ( \widetilde{\lambda}_{2}, \alpha ) ],
	\end{align}
where $\Gamma^{\prime}$ denotes the covariance matrix of $\sigma_{A}^{\prime} \otimes \sigma_{B}^{\prime}$, and $\widetilde{\lambda}_{1}$ and $\widetilde{\lambda}_{2}$ are the symplectic eigenvalues for $h ( h ( \Gamma^{\prime}, \Gamma ), \Gamma^{\prime} )$.

Finally, from the composition rule in Eq.~\eqref{eq:CPR}, the overlap between two Gaussian states having the same means is given by $[ \det ( \Gamma_{1} + \Gamma_{2} ) ]^{-1/2}$. The quantum Hilbert-Schmidt mutual information is then obtained as
	\begin{align}
		\mathcal{I}_{\mathrm{HS}}^{2} [ \sigma_{AB} ] = & \frac{1}{4\sqrt{(ab-c^{2})(ab-d^{2})}} + \frac{1}{4 ab} \nonumber \\
		- & \frac{2}{\sqrt{(4ab-c^{2})(4ab-d^{2})}}.
	\end{align}

\section{Covariance Matrices for Non-Gaussian States}
The covariance matrix of the entangled coherent state under a symmetric loss channel $\mathcal{L}_{\eta} [ \ketbra{\Psi}{\Psi} ]$ is given by
	\begin{equation}
		\Gamma_{\mathcal{L}_{\eta} [ \ketbra{\Psi}{\Psi} ]} =
			\begin{pmatrix}
				\mathcal{X} + \frac{1}{2} \mathbb{I}_{2}  & \mathcal{X} \\
				\mathcal{X} & \mathcal{X} + \frac{1}{2} \mathbb{I}_{2}
			\end{pmatrix},
	\end{equation}
where
	\begin{equation}
		\mathcal{X} = \frac{\eta \gamma^{2}}{\sinh 2 \gamma^{2}}
			\begin{pmatrix}
				e^{2 \gamma^{2}} & 0 \\
				0 & e^{- 2 \gamma^{2}}
			\end{pmatrix}.
	\end{equation}

For the case the photon number entangled state in the form $\sum_{k} c_{k} \ket{k}_{A} \ket{k}_{B}$, we obtain the corresponding covariance matrix as
	\begin{equation}
		\Gamma =
		\begin{pmatrix}
			a + \frac{1}{2} & & b & \\
			& a + \frac{1}{2} & & -b \\
			b & & a + \frac{1}{2} & \\
			& -b & & a + \frac{1}{2}
		\end{pmatrix},
	\end{equation}
with $a = \sum_{k} k |c_{k}|^{2}$ and $b= \sum_{k} (k+1) c_{k}^{*} c_{k+1}$.



\end{document}